\pdfoutput=1

\documentclass[reprint,aps,pre,showpacs,twocolumn,superscriptaddress,bibnotes,longbibliography]{revtex4-1}
\usepackage{microtype}
\usepackage{graphicx}
\usepackage{amsmath}
\usepackage{amssymb}
\usepackage[colorlinks,linkcolor=blue,urlcolor=blue,citecolor=blue]{hyperref}

\usepackage{mathptmx}
\usepackage{times}
\usepackage{epsfig,amsopn}
\usepackage{graphicx}
\usepackage{bm,amsmath,amssymb}
\usepackage{natbib}
\usepackage{braket}
\usepackage{float}
\usepackage{enumerate}
\usepackage{hyperref}
\usepackage[normalem]{ulem}
\allowdisplaybreaks[1]

\DeclareMathAlphabet{\mathcal}{OMS}{cmsy}{m}{n}

\begin{document}
\def\be{\begin{equation}}
\def\ee{\end{equation}}
\def\bea{\begin{eqnarray}}
\def\eea{\end{eqnarray}}
\def\f{\frac}
\def\l{\label}
\def\nn{\nonumber}

\definecolor{dgreen}{rgb}{0,0.7,0}
\def\redw#1{{\color{red} #1}}
\def\green#1{{\color{dgreen} #1}}
\def\blue#1{{\color{blue} #1}}
\def\brown#1{{\color{brown} #1}}

\newcommand{\eref}[1]{Eq.~(\ref{#1})}%
\newcommand{\Eref}[1]{Equation~(\ref{#1})}%
\newcommand{\fref}[1]{Fig.~\ref{#1}} %
\newcommand{\Fref}[1]{Figure~\ref{#1}}%
\newcommand{\sref}[1]{Sec.~\ref{#1}}%
\newcommand{\Sref}[1]{Section~\ref{#1}}%
\newcommand{\aref}[1]{Appendix~\ref{#1}}%
\newcommand{\sgn}[1]{\mathrm{sgn}({#1})}%
\newcommand{\erfc}{\mathrm{erfc}}%
\newcommand{\Erf}{\mathrm{erf}}%
%%%%%%%%%%%%%%%%%%%%%%%%%%%%%%%%%%%%%%%%%%%%%%%%%%%%%%%

\title{First passage under restart for discrete space and time: application to one dimensional confined lattice random walks}

\author{Ofek Lauber Bonomo and Arnab Pal}
\email{arnabpal@mail.tau.ac.il}

\affiliation{ School of Chemistry, Raymond and Beverly Sackler Faculty of Exact Sciences \& The Center for Physics and Chemistry of Living Systems \&   The Ratner Center for Single Molecule Science, Tel Aviv University, Tel Aviv 6997801, Israel}

\date{\today}

%%%%%%%%%%%%%%%%%%%%%%%%%%%%%%%%%%%%%%%%%%%%%%%%%%%%%
\begin{abstract}
{\noindent First passage under restart has recently emerged as a conceptual framework to study various stochastic processes under restart mechanism. Emanating from the canonical diffusion problem by Evans and Majumdar, restart has been shown to outperform the completion of many first passage processes which otherwise would take longer time to finish. However, most of the studies so far assumed continuous time
underlying first passage time processes and moreover considered continuous time resetting restricting out restart processes broken up into synchronized time steps. To bridge this gap, in this paper, we study discrete space and time first passage processes under discrete time resetting in a general set-up. We sketch out the steps to compute the moments and the probability density function which is often intractable in the continuous time restarted process. A criterion that dictates when restart remains beneficial is then derived. We apply our results to a symmetric and a biased random walker (RW) in one dimensional lattice confined within two absorbing boundaries. Numerical simulations are found to be in excellent agreement with the theoretical results. Our method can be useful to understand the effect of restart on the spatiotemporal dynamics of confined lattice random walks in arbitrary dimensions.}
\end{abstract}

\maketitle

%%%%%%%%%%%%%%%%%%%%%%%%%%%%%%%%%%%%%%%%%%%%%
\section{Introduction}
\label{Introduction}
There are certain benefits in starting anew. Restart or resetting, a new topic in statistical physics, teaches us that stopping intermittently and starting over again and again can increase the chances of reaching a desired outcome. Although it seems non-intuitive at a first glance, the basic physics is rather simple. Restart works by truncating the tails of long detrimental trajectories thus rendering the large stochastic fluctuations regular. In statistical physics, this mechanism was first observed in the canonical diffusion model with stochastic resetting by Evans and Majumdar \cite{Restart1,Restart2}. Since then, restart has emerged as a very active avenue of research in statistical physics \cite{Restart1,Restart2,Restart3,Restart4,Restart5,Restart6,PalJphysA} and generic stochastic process \cite{SP-0,SP-1,SP-2,SP-3,SP-4,SP-5,SP-6} due to its numerous applications spanning across interdisciplinary fields ranging from computer science \cite{Luby,algorithm}, population dynamics \cite{population-1}, queuing theory \cite{queue-1}, chemical and biological process \cite{ReuveniEnzyme1,bio-1,bio-2,bio-3}, foraging \cite{HRS} and search processes with rare events \cite{Montanari,extreme}. We refer to a recent review \cite{review} (and references therein) for a detailed account of the subject. The subject has also seen advances through single particle experiments using optical tweezers \cite{expt-1,expt-2}.

First passage under restart is one central direction in the field. This framework has been instrumental to study generic stochastic processes under various restart mechanisms \cite{PalReuveniPRL17}. The power of this approach lies on the fact that it allows one to compute general expressions for important metrics namely the moments and the distribution of completion time of restarted process regardless of the specifics of the underlying first passage process \cite{PalReuveniPRL17,ReuveniPRL16,branching,Belan,Chechkin,interval,Peclet,space}. Furthermore, it allows one to discover many universal phenomena such as a criterion for restart to be beneficial \cite{PalReuveniPRL17,Chechkin,potential-criterion-1,potential-criterion-2}, a globally dominating restart mechanism \cite{PalReuveniPRL17,Redner}, a general Landau like theory for restart transitions \cite{interval-v} and conditions on other quantiles of first passage processes \cite{quantile} that emerge as an effect of restart.

\begin{figure}[b]
\includegraphics[scale=0.65]{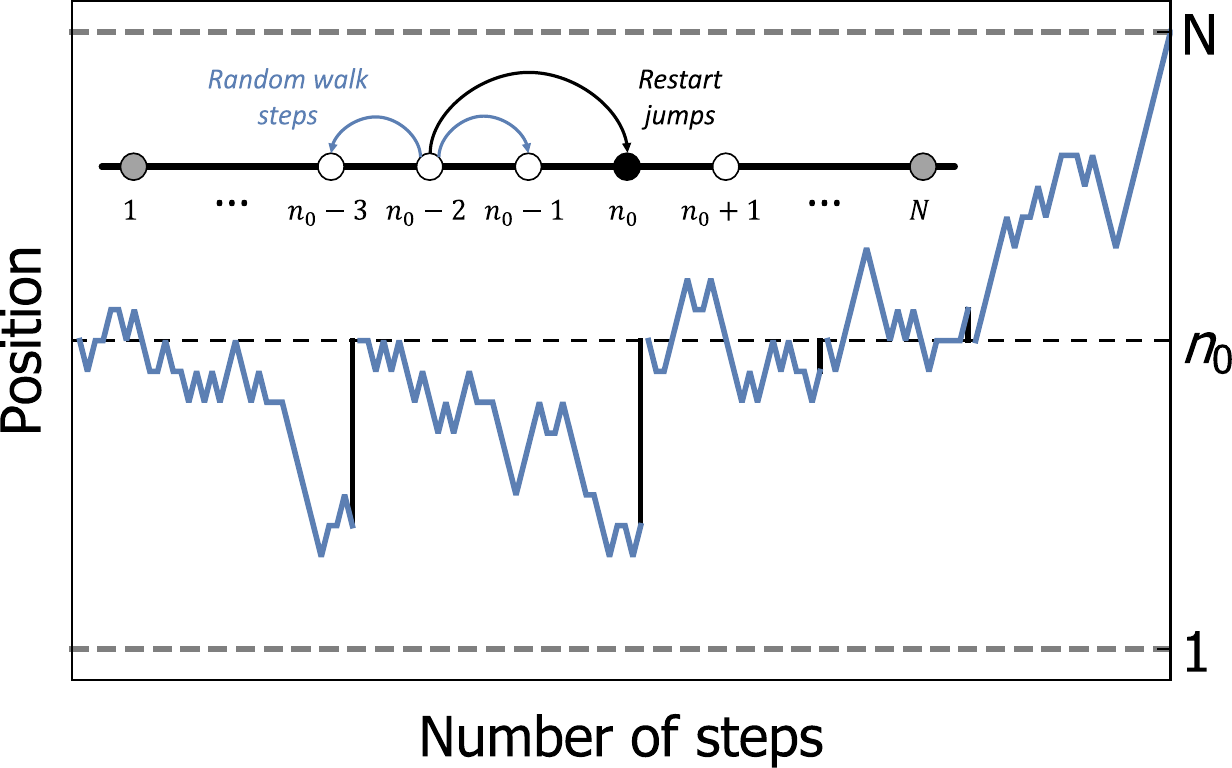}
\caption{Schematic of a lattice random walker in a 1D confined geometry under discrete restart. First passage occurs as soon as the walker reaches one of the boundaries located at $1$ and $N$. The restart coordinate is $n_0$, same as the initial condition.}
\label{Schematic}
\end{figure}
 
A Brownian walker on a real line subject to a resetting to the origin at certain rate is perhaps the most quintessential example of restart phenomena \cite{Restart1}. In this problem, one is interested in the first passage time to a target which is located at a given distance and it was shown that restart can expedite the completion.  This problem was then studied in higher dimensions \cite{high-D} and complex geometries \cite{bounded,comb}, in the presence of multiple targets \cite{interval,bressloff} and also in the presence of generally distributed resetting time density \cite{PalJphysA,PalReuveniPRL17,Chechkin}. It is important to emphasize that resetting occurring at a constant rate essentially implies that the waiting times between resetting events are taken from an exponential distribution. In other words, here, resetting is a continuous time Markov process. Notably, majority of the studies in the field spanned around the continuous time Brownian motion and its variants e.g., scaled diffusion \cite{scaled}, underdamped \cite{underdamped} and random acceleration process \cite{RAP} (also see \cite{review} for other model systems). A general version of random walk namely the mesoscopic continuous time random walks with heavy tailed jump distributions (such as L\'evy flights) were studied in \cite{CTRW-0,CTRW-1}. Continuous-time RWs have also been studied under various resetting strategies (power law, Markovian etc.) \cite{CTRW-2,CTRW-3,CTRW-4,CTRW-5}.
In a recent work \cite{Christophorov}, a continuous time lattice RW was considered in the presence of resetting conducted at a rate and stationary probability distribution, mean first passage time were computed for an infinite and semi-infinite chain. Interacting continuous time lattice RWs with exclusion were studied in the presence of stochastic resetting \cite{SEP,local}. Continuous time RWs on a network in the presence of stochastic resetting were studied in \cite{network-1,network-2}. Arguably, most of these models also use resetting to be a continuous time process for the computations there have neater mathematical structure.

Nonetheless, there are some recent works which shed light on the discrete time lattice RW with discrete resetting. A discrete time RW on a real line (but with jumps drawn from a continuous and symmetric distribution) in the presence of resetting was studied in \cite{SP-0}. A discrete time unidirectional RW namely 
Sisyphus lattice RW was studied in the presence of the resetting
probability which can be random or site-dependent \cite{SisyphusRW}. They analyzed the first passage time and survival probabilities for the walker to reach a certain threshold in the lattice. Recently, preferential visit models have been introduced to generalize resetting with memory of history \cite{Boyer1}. This is most naturally illustrated by a discrete time lattice RW which with some reset probability is returned to its previous position at a randomly selected time from the past. More precisely, the walker relocates to a previously
visited site with a probability proportional to the number of past visits to that site. It was shown that the RWs perform slow sub-diffusion due to the dynamics of memory-driven resetting \cite{Boyer1,Boyer2}. The model above was further studied in the presence of a single defect site where the RW stays with a finite probability \cite{Boyer3,Boyer4}. Another discrete time RW on a lattice was considered in \cite{maximum} where resetting with a probability relocated the walker to the previous maximum. It becomes quite evident that unlike the continuous time and space first passage processes under continuous time resetting, the discrete counterparts are only handful. Moreover, they are studied keeping the specifics of the problem in hand. We aim to bridge this gap by building a general framework for discrete first passage processes under discrete time resetting. To this end, we follow the steps of \cite{PalReuveniPRL17} developed by one of the authors of the current paper. Importantly, this approach holds even when both the first passage and restart processes are not necessarily Markovian as long as the memory from the past is erased after each resetting. Pertaining to this property, renewal framework has been used extensively in the above-mentioned fields along with in stochastic thermodynamics \cite{sth}, quantum mechanics \cite{quantum-1} and nonlinear dynamics \cite{dynamical}. We provide working formulas for the moments and the probability density function for the first passage time under restart. We then derive a condition which asserts when restart is going to expedite the completion. We employ our results to the application of 1D lattice RWs in confined geometry under two different restart strategies.

Lattice RWs are a special class of Markov processes which were popularized following P\'{o}lya’s seminal work on the dimensionality dependence of the recurrence probability, that is, the probability that a random walker on an
\textit{infinite space} $d-$dimensional lattice eventually returns to its starting point \cite{LRW-1,LRW-2,LRW-3,LRW-4,LRW-5,LRW-6}. Needless to say that the subject is now a text book material and the applications are myriad. Somewhat surprisingly, and in contrast to Brownian walks, the space-time dependence of the \textit{confined lattice walk} probability has been accessible mainly via computational techniques due to combinatorial hindrances and only a few exact results exist. Although for 1D confined domains, the time-dependent propagator for absorbing boundaries \cite{Feller} and periodic domains \cite{periodic} were known, no analytical results were known for mixed or reflective boundary conditions in 1D and in high dimensions for all the above-mentioned boundary conditions. Only recently, Giuggioli has derived exact results for the space and time dependence of the occupation
probability, first passage time probability for confined P\'{o}lya’s walks in arbitrary dimensions with reflective, periodic, absorbing, and
mixed (reflective and absorbing) boundary conditions along each direction \cite{Luca1}. Further generalizations were made by Sarvaharman and Giuggioli for the biased RWs with different boundary conditions \cite{Luca2}. Our derivation shows that the solution for the restarted process can be given in terms of the solution for the underlying process and thus reducing the overall complexity in manifolds. Naturally, these very recent results by Giuggioli and coauthors set the perfect stage for us to employ them directly when restart is involved.

The paper is organized as follows. In \sref{FPUR}, we build up the framework of first passage step under restart, derive the working formulas for the statistical moments, and the probability mass function for the completion time. In the consecutive subsections, we discuss two different restart strategies namely the geometric and sharp protocols. We herein derive the sufficient criterion for geometric restart to be beneficial for any first passage process. Next in \sref{applications}, we apply the framework to one dimensional lattice RWs. In \sref{example-1}, we discuss the simple random walk while in \sref{example-2} we present the biased random walk in the presence of the above-mentioned restart strategies.  Our conclusions are summarized in \sref{conclusions}. Some of the derivations from the main text and additional discussions are reserved for the Appendix. Before we proceed, it will be useful to introduce the notations used throughout the paper. We will use $P_X(x)$, $\langle X \rangle$, $\sigma_X ^2$, and $G_X(z) \equiv \langle z^X \rangle$ to denote, respectively, the probability mass/density function (PMF/PDF), mean/expectation, variance, and the probability generating function (PGF) of a discrete random variable $X$ taking values in the non-negative integers.

\section{First passage step under restart}
\label{FPUR}
Consider a generic discrete step first passage process that starts at the origin and, if allowed to take place without interruptions, ends after a random number of steps $N$. The process is,
however, restarted after some random number of steps $R$. Thus, if the process is completed prior, or at the same time as the restart, we mark a completion of the event. Otherwise, the process will start from scratch and begin completely anew. This procedure repeats itself until the process reaches completion. Denoting the random completion number of steps of the restarted process by $N_R$, it can be seen that
\bea
N_{R}=\begin{cases}
N & N \leq R\\
R+N'_{R} & N> R, \label{Renewal eq}
\end{cases}
\eea 
where $N'_{R}$ is an independent and identically distributed copy of $N_{R}$. \eref{Renewal eq} is the central renewal equation for first passage step under restart and assumes that after each restart, the memory is erased from the previous trial. To obtain the mean number of steps for the restarted process, we note that \eref{Renewal eq} can be written as $N_R=min\left( N,R \right)+I\left\{ N>R\right\} N'_{R}$, where $I\left\{ N>R\right\}$ is an indicator random variable that takes the value 1 when $N>R$ and zero otherwise. Taking expectations on the both sides and using that $N$ and $R$ are independent of each other, we find the \textit{mean completion time} under restart to be
\bea 
\left\langle N_{R}\right\rangle =\frac{\left\langle min\left(N,R\right)\right\rangle }{\text{Pr}\left(N\leq R\right)}. \label{mean FPUR}
\eea 
The numerator can be computed by noting that the probability $\text{Pr}\left(min\left(N,R\right)>n\right)=\text{Pr}\left(N>n\right)\text{Pr}\left(R>n\right)$ and thus
\bea 
\left\langle min\left(N,R\right)\right\rangle&=&\sum_{n=0}^{\infty}\text{Pr}\left(min\left(N,R\right)>n\right) \nn \\
&=& \sum_{n=0}^{\infty} \left(\sum_{k=n+1}^{\infty}P_N\left(k\right)\right) \left(\sum_{m=n+1}^{\infty}P_R\left(m\right)\right), \label{Mean min formula}
\eea 
where recall that $P_{N}(n)$ and $P_{R}(n)$ are the probability density functions for the first passage and restart process respectively. On the other hand, the denominator in \eref{mean FPUR} can also be computed easily 
\bea
\text{Pr}\left(N\leq R\right)=\sum_{n=0}^{\infty}P_{N}(n)\sum_{m=n}^{\infty}P_{R}(m).  \label{denominator}
\eea
We now turn our attention to derive the generating function for the restarted process. The probability generating function of the discrete random variable $N_R$ taking values in the non-negative integers ${0,1, ...}$, $n$, is defined as
\bea 
G_{N_R}(z) \equiv \left\langle z^{N_R}\right\rangle = \sum_{n=0}^{\infty} P_{N_R}(n)z^{n}, \label{PGF def}
\eea 
where $P_{N_R}(n)$ is the probability mass function of $N_R$. It will now prove useful to introduce the following conditional random variables
\bea 
N_{min}&=&\left\{ N_{R}|N \leq R\right\} =\left\{ N|N \leq R\right\} %=\left\{ N|N=min\left(R,N\right)\right\}
, \label{Tmin main} \\
R_{min}&=&\left\{ N_R|N> R\right\} =\left\{ R|N> R\right\} %=\left\{ R|R=min\left(R,N\right)\right\}
, \label{Rmin main}
\eea 
with their respective densities
\bea 
P_{N_{min}}(n)&=&P_{N}(n)\frac{\sum_{m=n}^{\infty}P_{R}(m)}{\text{Pr}\left(N \leq R\right)}, \label{Nmin mass-main} \\
P_{R_{min}}(n)&=&P_{R}(n)\frac{\sum_{m=n+1}^{\infty}P_{N}(m)}{\text{Pr}\left(N > R\right)}, \label{Rmin mass-main}
\eea 
where $\text{Pr}\left(N > R\right)=1-\text{Pr}\left(N \leq R\right)$.
Using the renewal \eref{Renewal eq} and the new random variables in Eqs.(\ref{Tmin main})-(\ref{Rmin main}), we can write (see \aref{derive-G}) 
\begin{align}
    G_{N_{R}}(z)&=&\text{Pr}\left(N \leq R\right)\left\langle z^{N_{min}}\right\rangle +\text{Pr}\left(N > R\right)\left\langle z^{R_{min}+N'_{R}}\right\rangle.
\end{align}
Now using the fact that $N'_{R}$ is an independent
and identically distributed copy of $N_{R}$ in above, we arrive at the following expression for the generating function of the restarted process
\bea 
G_{N_{R}}(z)
&=&\frac{\left(1-\text{Pr}\left(N > R\right) \right)G_{N_{min}}(z)}{1-\text{Pr}\left(N > R\right)G_{R_{min}}(z)}, \label{PGF final result}
\eea 
where $G_{X_{min}}(z)$ is the generating function for the random variable $X_{min}$. The above formula (\ref{PGF final result}) is extremely useful since it allows one to compute \textit{all the moments}
\bea 
\left\langle N_R^{k}\right\rangle &=&\left(z \frac{\partial}{\partial z}\right)^{k}G_{N_R}(z)\Big|_{z=1^{-}}, \label{PGF moments}
\eea 
and, importantly, also the \textit{probability density function} of $N_R$
\bea
P_{N_R}(n)&=&\text{Pr}(N_R=n)=\frac{G_{N_R}^{(n)}(0)}{n!}, \label{PGF PMF}
\eea
where $\left\langle N_R^{k}\right\rangle$ is the $k$-th moment and $G_{X}^{(n)}(z)$ is the $n$-th derivative of $G_{X}(z)$ with respect to $z$, and $n=0,1,2...$. It is important to emphasize that in continuous time set-ups, deriving the first passage time density for the restarted process requires Laplace inversions, and thus it remains intractable in most of the cases (except some asymptotic limits \cite{Restart1}). In stark contrast, in discrete set-up, computation of the first passage time density requires only derivatives of the generating function as seen in \eref{PGF PMF}, and thus is more accessible. It is easy to see that the expression for the mean in \eref{mean FPUR} can be easily recovered using the generating function given in \eref{PGF final result} and noting
\begin{align}
\langle N_{R}\rangle &= z \frac{\partial G_{N_R}(z)}{\partial z} \Big|_{z=1^{-}} \nn \\
&= \scriptstyle{\frac{z\left(1-\text{Pr}\left(N> R\right)\right)\left(1-\text{Pr}\left(N> R\right)G_{R_{min}}\left(z\right)\right)G'_{N_{min}}\left(z\right)-\text{Pr}\left(N> R\right)G_{N_{min}}\left(z\right)G'_{R_{min}}\left(z\right)}{\left(1-\text{Pr}\left(N> R\right)G_{R_{min}}\left(z\right)\right)^{2}}\Big|_{z=1^{-}}}. 
\label{Mean number of steps 1}
\end{align}
Substituting $z\rightarrow1^-$, and utilizing the relations $G_{N_{min}}(1^-)=G_{R_{min}}(1^-)=1$, $G'_{N_{min}}(1^-)=\langle N_{min} \rangle, G'_{R_{min}}(1^-)=\langle R_{min} \rangle$, obtained from Eqs. (\ref{Nmin mass-main})-(\ref{Rmin mass-main}), we get
\begin{align}
\langle N_{R}\rangle
&=\frac{\text{Pr}\left(N\leq R\right)\left\langle N_{min}\right\rangle +\text{Pr}\left(N > R\right)\left\langle R_{min}\right\rangle }{\text{Pr}\left(N \leq R\right)} \nn \\
&= \frac{\langle min\left(N,R\right) \rangle}{\text{Pr}\left(N \leq R\right)},
\label{Mean number of steps 1}
\end{align}
which is indeed \eref{mean FPUR}. Similarly, the \textit{second moment} can be computed using the following relation 
\bea
\langle N^2_R \rangle
&=& z G'_{N_R}(z)+z^2G''_{N_R}(z) \Big|_{z=1^{-}} .\label{second moment FPUR}
\eea 
Higher order moments can also be computed in a similar way.
So far, we have kept our formalism extremely general without specifying the forms of the restart time density. In what follows, we will use 
two different distributions for the restart time namely the geometric and the sharp distribution.

\subsection{Geometrically distributed restart}
Consider a resetting number step taken from a geometric distribution with parameter $p~  (0<p<1)$, 
\bea
P_R(n)=(1-p)^{n} p, ~~n\geq 0.
\eea
In other words, restart would take place with a probability $p$ after $n$ unsuccessful trials. Notably, this distribution is the discrete analog of the exponential distribution, being the only discrete distribution possessing the memory-less property \cite{Mor}. For this distribution, following \aref{derivation-geometric-appendix}, we have
\bea
\left\langle min\left(N,R\right)\right\rangle&=&\frac {1 - p} {p}\left(1-G_N(1-p)  \right), \label{mean min geometric-main}\\
\text{Pr}\left(N\leq R\right)&=&G_N(1-p). \label{Mean denominator-main}
\eea

Substituting Eqs. (\ref{mean min geometric-main}) and (\ref{Mean denominator-main}) into the formula for the mean of the restarted process given in \eref{mean FPUR} yields 
\bea 
\langle N_R \rangle &=& \frac{1-G_N(1-p)}{G_N(1-p)} \frac {1 - p} {p}.
\label{mean FPUR geometric}
\eea 
The above expression can be understood intuitively --- while the first fraction gives the mean number of restart events till the first passage occurs, the second fraction simply quantifies the mean number of steps taken between any two restart events.
At the limit $p\rightarrow 0^+$, namely, when restart is rare, \eref{mean FPUR geometric} reduces to (following L'Hospital's rule) 
$\lim_{p\to 0^+} \langle N_R \rangle=G'_N(1^-) = \langle N \rangle$.

We now turn to the derivation of the PGF of the restarted process under geometric restart. We first compute the PGFs for the conditional random variables $N_{min}$ and $R_{min}$. Following \aref{derivation-geometric-appendix}, we find
\bea
G_{N_{min}}(z)&=&\frac{G_N(z(1-p))}{G_N(1-p)}, \nonumber \\
G_{R_{min}}(z)&=&\frac{p\left( G_N(z(1-p))-1 \right)}{((1-p)
   z-1) \left(1-G_N(1-p)\right)}.
\label{GNmin-Rmin}
\eea
Substituting \eref{GNmin-Rmin} into \eref{PGF final result} we find
\bea
G_{N_{R}}(z)
&=& \frac{(1-(1-p) z) G_N((1-p) z)}{(1-p)
   (1-z)+p G_N((1-p) z)}, \label{PGF geometric}
\eea 
from which one can derive the probability mass function of the restarted process by taking the derivatives of the generating function using \eref{PGF PMF}. We compute the second moment using the above expression in \eref{second moment FPUR}  
\bea
\langle N^2_R \rangle
&=& \scriptstyle{\frac{(1-p) \left(G_N(1-p) (3p-2-pG_N(1-p))+\left(2 p^2-2 p\right)
   G'_N(1-p)-2 p+2\right)}{p^2
   G^2_N(1-p)}}. \label{second moment FPUR geometric}
\eea 
 One of the hallmark properties of the restart is its ability to lower the underlying mean first passage time and this often leads to an optimal value of the restart rate at which the mean time reaches a global minimum. To elucidate this in the discrete set up, we now study whether there exists a sufficient enough criterion under which restart is always beneficial.

\subsection{A criterion for geometric restart to be beneficial}
\label{CV-criterion-geometric}
To derive the criterion, we first observe a first passage time process and turn on an infinitesimal restart probability $p \to 0^+$. If restart has to lower the mean time, it is sufficient enough to check whether $d\langle N_R \rangle/dp|_{p \to 0}<0$, where $\langle N_R \rangle$ is given by \eref{mean FPUR geometric}. A small $p$ expansion of $\langle N_R \rangle$ gives
\bea 
\langle N_R \rangle \approx G_N'(1)+\frac{1}{2} p \left(2 G_N'(1)^2-2 G_N'(1)-G_N''(1)\right). \label{Mean around p = 0}
\eea 
Now noting that $G_N'(1)=\langle N \rangle$, $G_N''(1)+G_N'(1)-G_N'(1)^2=Var(N)$ and substituting into \eref{Mean around p = 0}, the criterion can be recast as
\bea
CV^2>1-\frac{1}{\langle N \rangle}
\label{CV crierion}
\eea
where $CV^2=\frac{Var(N)}{\langle N \rangle^2}$ is the squared coefficient of variation of the underlying first passage process. This essentially means that whether restart would favour a completion depends on the underlying first passage time process. Moreover, this criterion is also not sensitive to the entire density, but only to the first two moments of the underlying process. We refer to a similar criterion that was derived for the continuous stochastic resetting case in \cite{PalReuveniPRL17}.

It is, however, clear that repeated restart will only prolong the completion since the process is almost `frozen' to the resetting configuration, and thus the average completion time will be exceedingly large. Taking this fact along with the criterion \eref{CV crierion} simply implies that the mean completion time must be having at-least one minimum as a function of $p$.
The optimal probability, $p^*$, which minimizes this mean first passage time
(\eref{mean FPUR geometric}), can be determined from the following root equation
\begin{align}
G^2_N(1 - p^*) - G_N(1 - p^*) + (1-p^*) p^*G'_N(1 - p^*) = 0. \label{optimal p}
\end{align}
Clearly, the valid solution of the above equation provides us with an optimal probability $p^*$ for which the mean completion time attains a minimum and thus adheres to the criterion (\ref{CV crierion}).

\subsection{Sharp restart}
We also consider a strategy when restart events always take place after a fixed number of steps. This is often known as sharp or deterministic restart protocol (see \cite{PalJphysA,PalReuveniPRL17,sharp-Reuveni} for different properties of this strategy in the continuous set-up) and the density is given by
\bea 
P_R(n)=\delta_{n,r}=\begin{cases}
0, & n\neq r\\
1, & n=r
\end{cases}, \label{Sharp distribution}
\eea 
where $\delta_{n,r}$ is the Kronecker delta. So, we will refer to this as sharp distribution with restart step $r$. 
For sharp restart, we have $\text{Pr}\left(N\leq R\right)=\sum_{n=0}^{r}P_{N}(n)=\text{Pr}\left(N\leq r\right).$
Furthermore, we find (\aref{Sharp-numerics})
\bea
G_{N_{min}}(z)&=&\frac{1}{\text{Pr}\left(N \leq r\right)} \sum_{n=0}^{r} P_{N}(n) z^n,\\
G_{R_{min}}(z)&=&z^r.
\label{G-rmin-nmin-z}
\eea
Substituting Eq. (\ref{G-rmin-nmin-z}) into \eref{PGF final result} we get 
\bea
G_{N_{R}}(z)
&=& \frac{\sum_{n=0}^{r} P_{N}(n) z^n}{1-\text{Pr}\left(N > r\right)z^r}. \label{PGF sharp-main}
\eea 
Again, using \eref{PGF PMF} one can derive the full probability mass function of the restarted process by taking the derivatives of the PGF given in \eref{PGF sharp-main}.
The moments can be computed using \eref{PGF moments} e.g., the first and second moment take the following forms
\bea 
\langle N_R \rangle = \frac{\sum _{n=0}^r n P_N(n)}{\text{Pr}\left(N \leq r\right)}+\frac{\left(1-\text{Pr}\left(N \leq r\right) \right)r}{\text{Pr}\left(N \leq r\right)}, \label{mean sharp}
\eea 
and
\begin{align}
\langle N^2_R \rangle =  
\scriptstyle{\frac{(\text{Pr}\left(N > r\right) (2 r-1)+1) \sum _{n=0}^r n P_N(n)+\text{Pr}\left(N \leq r\right) \sum _{n=0}^r (n-1) n
   P_N(n)+\text{Pr}\left(N > r\right) (1+\text{Pr}\left(N > r\right)) r^2}{\text{Pr}\left(N \leq r\right)^2}}. \label{Second moment sharp}
\end{align}
Here only when $r \gg 1$ (i.e., when restart events are rare), we reach to the limit of the underlying process e.g., $\langle N_R \rangle \to \langle N \rangle$ etc. 

\begin{figure}[t]
\includegraphics[width=8.5cm,height=2.8cm]{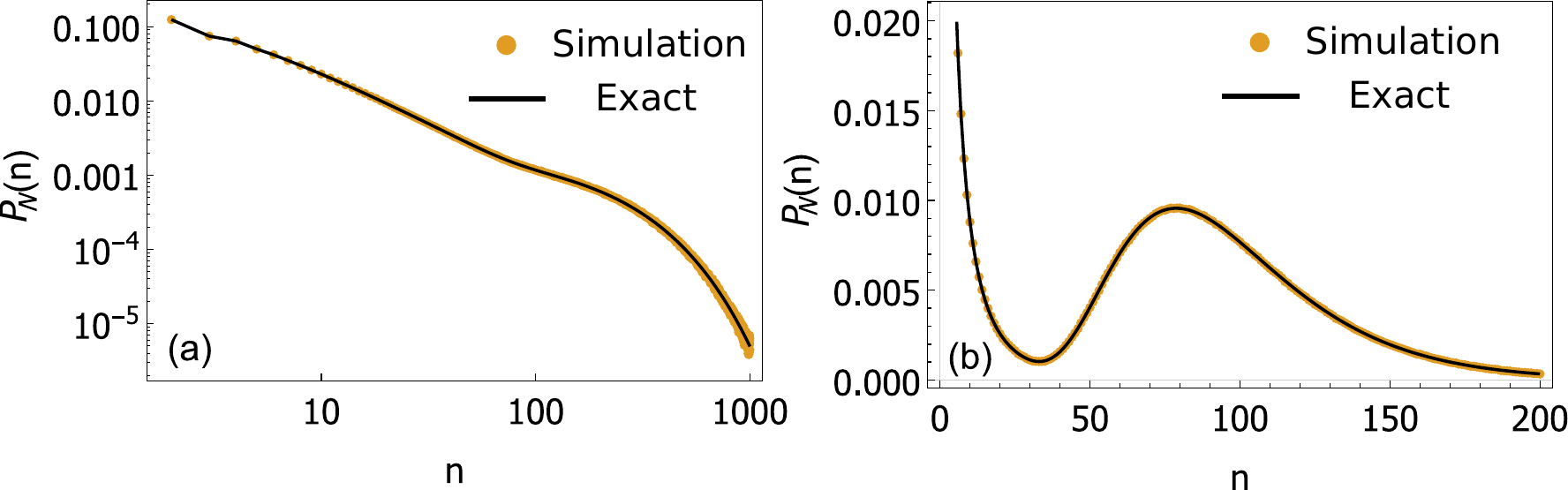}
\caption{First passage time densities $P_N(n)$ for the underlying process. Analytical results are taken from Eqs. (\ref{Sym RW underlying}), (\ref{FP from survival}) and plotted against numerical simulations. Panel (a): symmetric RW with diffusivity parameter $q=0.7$. Panel (b): biased RW with $q = 0.7$ and bias $g = 0.3$. Other parameters are kept fixed for both the simulations: left boundary $ = 1$, right boundary $ = 25$ and initial position $n_0 = 3$.}
\label{Underlying-FPT}
\end{figure}

\section{Applications to Random Walks}
\label{applications}
So far, we have built up a general framework for first passage under restart in discrete space and time. We have provided working formulas to compute the mean, higher moments, and even the probability mass function for the completion time without specifying any details of the underlying process. To see how they work in practice, we apply the formalism to a symmetric and biased random walk in a 1D lattice in the presence of two absorbing boundaries. The dynamics is further subjected to restart which after a random step brings the walker back to its initial state. We start with the symmetric unbiased random walk.

\subsection{Symmetric random walk in 1D}
\label{example-1}
\begin{figure}[b]
    \includegraphics[width=8.5cm,height=6.cm]{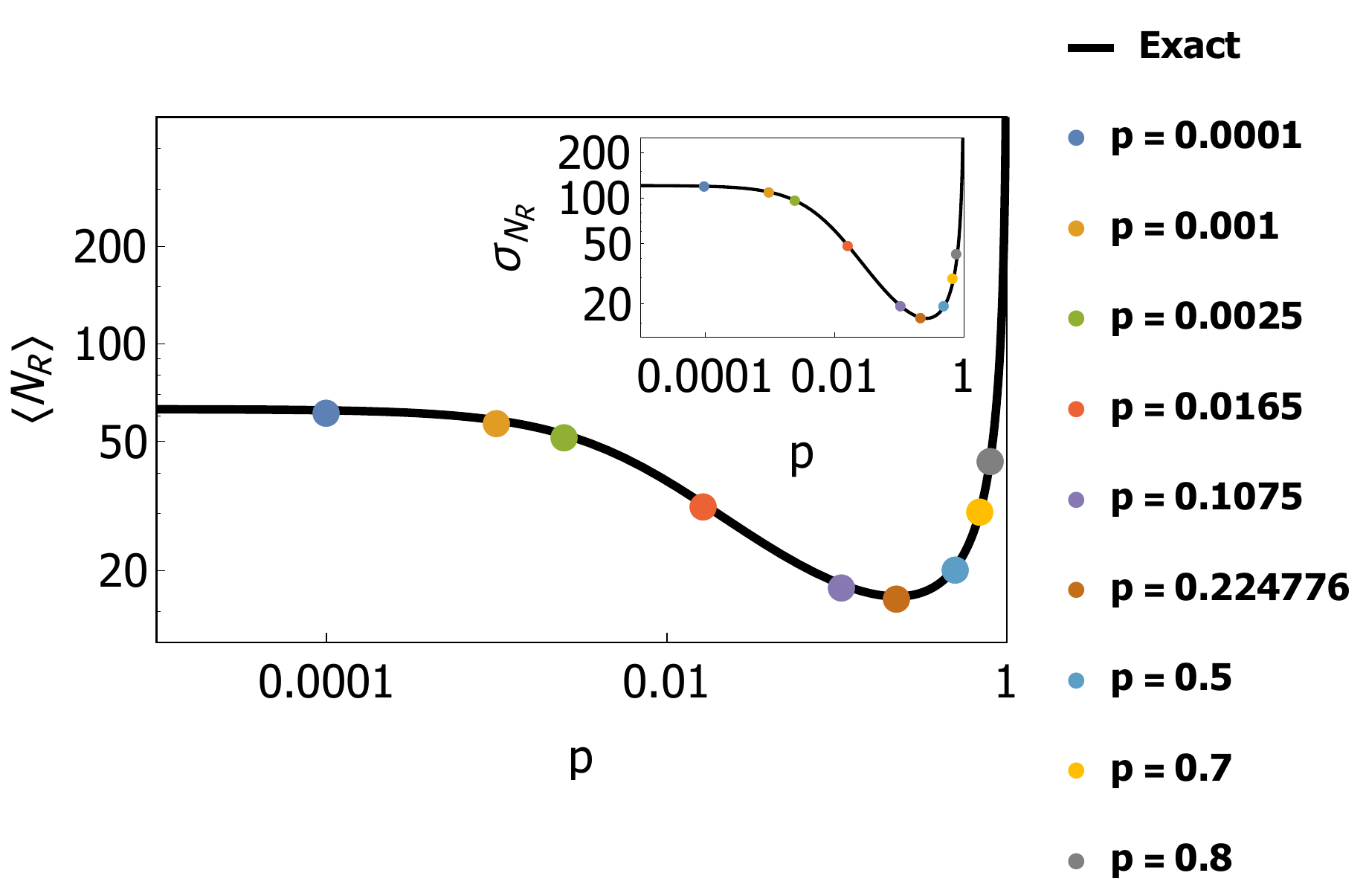}
    \caption{Mean first passage time $\langle N_R \rangle$ and fluctuations $\sigma_{N_R}$ (inset) of a symmetric RW in confinement as a function of restart probability $p$. Simulations are indicated by the markers while the continuous line represents the exact result. For $p \to 0^+$, these statistical quantities saturate to their underlying values. As can be seen, for a range of restart probabilities, mean time could be significantly reduced.
    Parameters set for the simulations: left boundary $ = 1$, right boundary $ = 25$, initial position $n_0 = 3$ and diffusivity parameter $q = 0.7$.}
    \label{Non biased RW}
\end{figure}

\begin{figure*}[t]
\includegraphics[width=8.2cm,height=4.3cm]{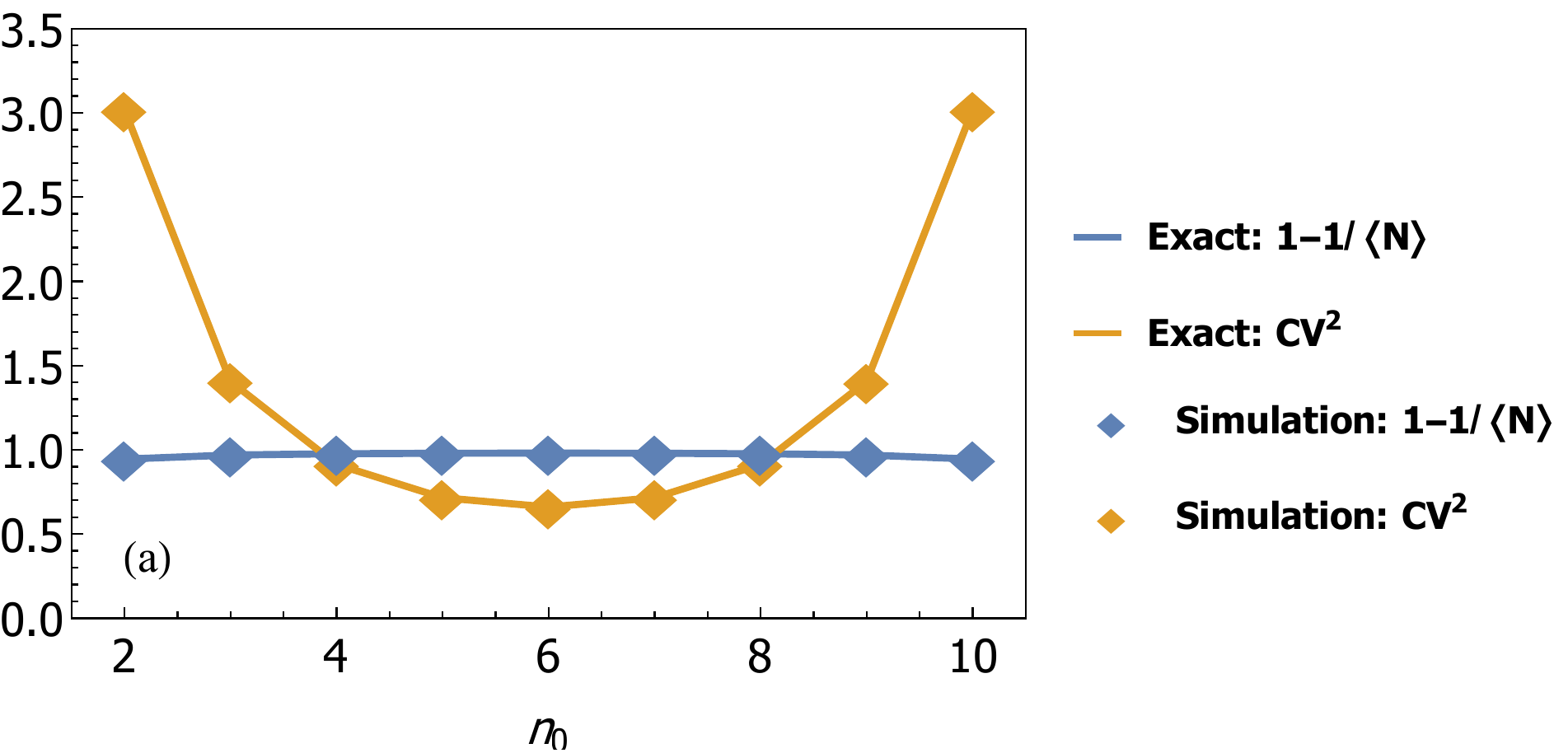}
\includegraphics[width=8.2cm,height=4.2cm]{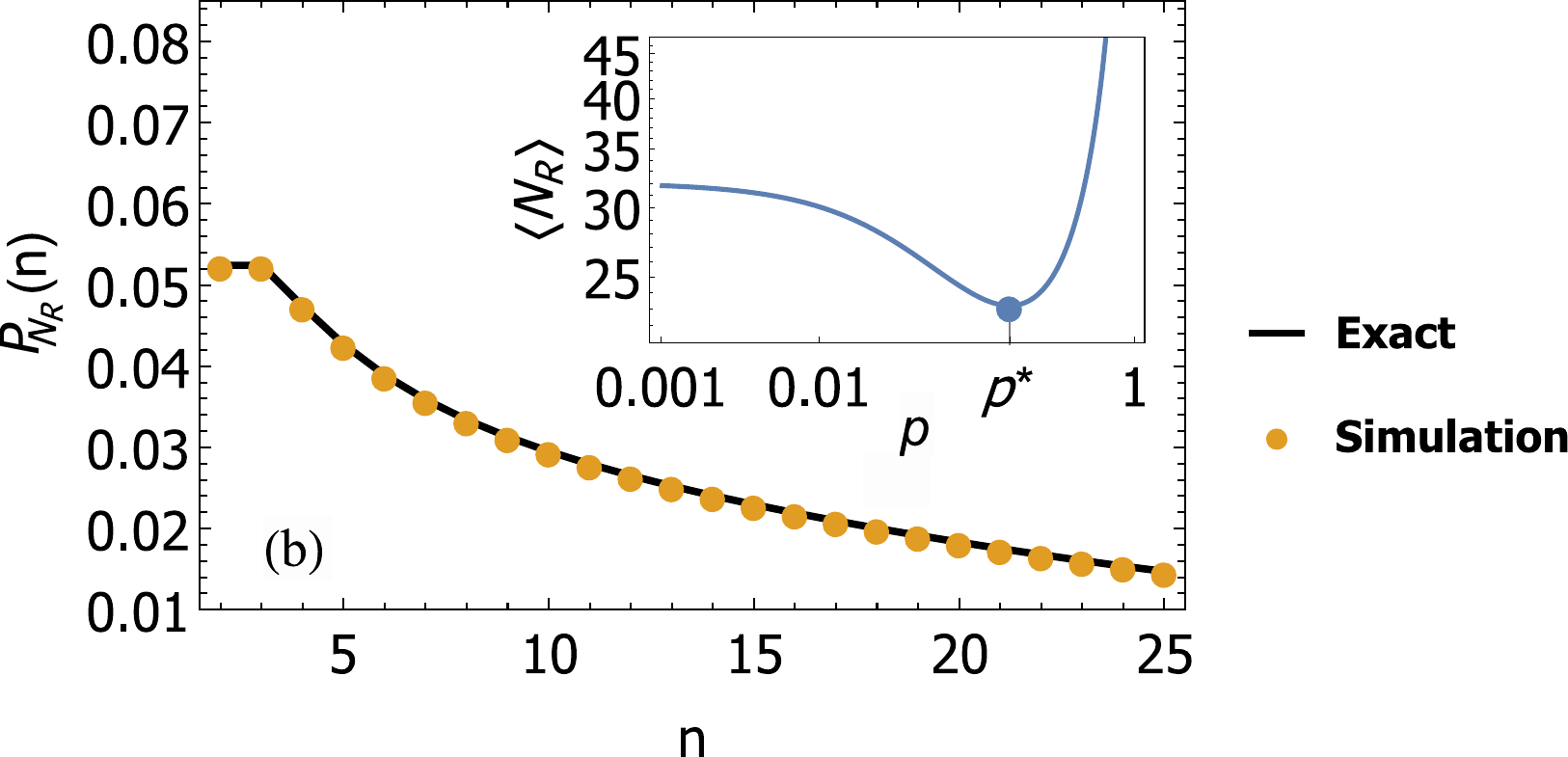}
\caption{Panel (a): demonstration of the restart criterion (\ref{CV crierion}) for the symmetric RW in confinement. We plotted the LHS and RHS of the criterion as a function of the initial position $n_0$. Panel (b): we choose $n_0=3$ from the panel (a) for which restart is beneficial, and then plot the probability mass function $P_{N_R}(n)$ for this given parameter value. In the inset, we plot $\langle N_R \rangle$ to demonstrate that indeed restart is able to reduce the mean completion time. The optimal probability $p^*=0.1610..$ is found to match exactly with the theoretical prediction. Parameters fixed for the simulations: left boundary $ = 1$, right boundary $ = 11$ and $q = 0.5$.}
\label{Non biased RW-criterion}
\end{figure*}

\begin{figure}[b]
\includegraphics[width=8.5cm,height=3.25cm]{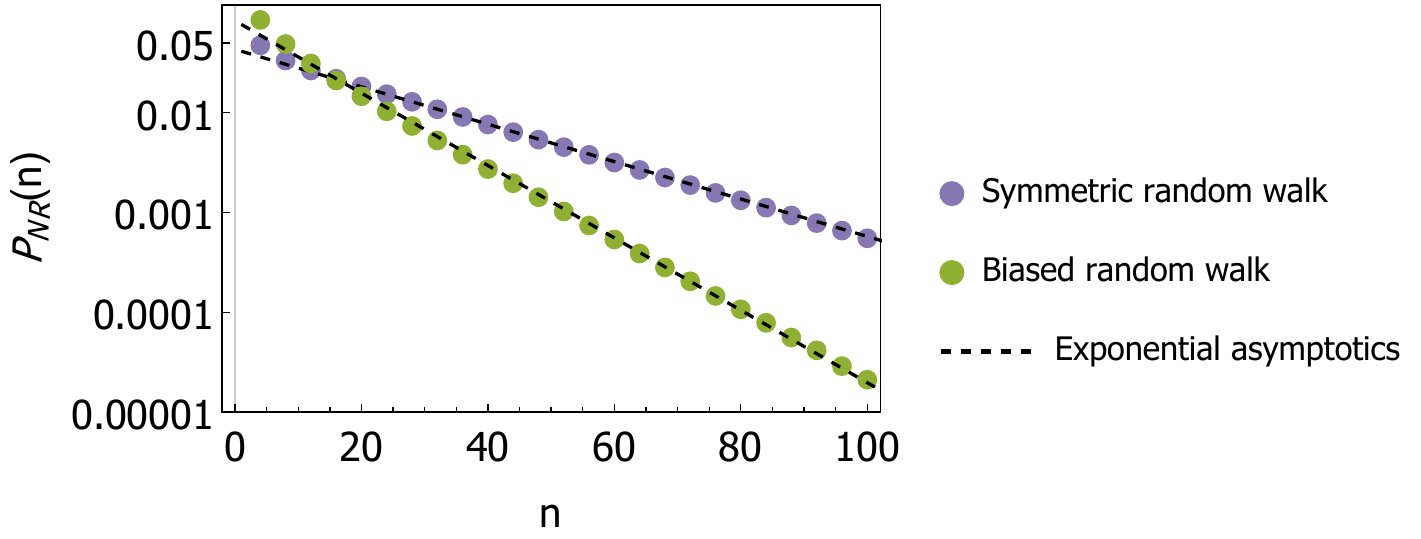}
\caption{Comparison between $P_{N_R}(n)$ and numerical data. The asymptotic forms of $P_{N_R}(n) \sim e^{-n/\langle N_R \rangle}$ are taken along with their respective $\langle N_R \rangle$ for the symmetric and biased RW in confinement. Parameters fixed for the symmetric RW: left boundary $ = 1$, right boundary $ = 11$, initial position $n_0 = 3$, $q = 0.5$ and $p=0.1610..~$. Parameters fixed for the biased RW: left boundary $ = 1$, right boundary $ = 11$, initial position $n_0 = 9$, $q = 0.5,~g = 0.3$ and $p=0.0813..$.}
\label{scaling}
\end{figure}

Let us consider the so-called symmetric lazy random walker \cite{Luca1, Lazy walker}, in 1D bounded domain. The dynamics of this walker are governed by the following evolution equation for the site occupation probability, $\mathbb{P}(m,n)$
\begin{align}
\mathbb{P}(m, n+1) =(1 - q) \mathbb{P}(m, n) + \frac{q}{2} \mathbb{P}(m - 1, n)  + \frac{q}{2}\mathbb{P}(m + 1, n),
\label{no bias mater eq}
\end{align}
with $m$ representing a lattice site on the line and $n$ being
time step. The $q$ parameter for this kind of random walk represents the tendency of the walker to move, with $q = 0$ representing a walker that does not move, and $q = 1$ a walker that moves at each time step. The walk is symmetric since the probability to go either to the left or to the right is $q/2$. The walker sets off from $m=n_0$ and stays inside the bounded interval $(1,N)$, with absorbing boundaries located at $m = 1$ and $m = N$ so that $\mathbb{P}(1,n)=\mathbb{P}(N,n)=0$. The first-passage time probability density i.e., the time for the walker to reach either of the two boundaries, $P_N(n)$, is given by \cite{Luca1}
\bea 
&&P_N(n)= \nn \\
&&\scriptstyle{\frac{q}{N-1}\sum_{k=1}^{N-2}\left[1-\left(-1\right)^{k}\right]\sin\left[\frac{n_{0}-1}{N-1}\pi k\right]\sin\left(\frac{\pi k}{N-1}\right)\left[1-q+q\cos\left(\frac{\pi k}{N-1}\right)\right]^{n-1}}, \label{Sym RW underlying}
\eea 
also see \fref{Underlying-FPT}a for a numerical verification. The generating function of the first passage time density can be obtained by noting $G_N(z)=\sum_{n=0}^\infty~z^nP_N(n)$, where $P_N(n)$ is given by \eref{Sym RW underlying}. 
 Substituting the resulting expression for $G_N(z)$ into \eref{mean FPUR geometric} with $z \to 1-p$ yields the mean first passage time under geometric restart. This is demonstrated in \fref{Non biased RW}. Moreover, from the generating function (\ref{mean FPUR geometric}) we obtain the second moment using \eref{second moment FPUR geometric}, and then plot the fluctuations $\sigma_{N_R}$ of first passage time as a function of $p$ as shown in the inset of \fref{Non biased RW}.
It is clear from the \fref{Non biased RW} that restart aids in the favour of completion. In other words, this must satisfy the restart criterion given in \eref{CV crierion}. The criterion is illustrated in \fref{Non biased RW-criterion}a where we have plotted the LHS and RHS of the inequality (\ref{CV crierion}) keeping $n_0$, the initial position/restart location, as the controlling parameter. We pick $n_0=3$ for which $CV^2>1-\frac{1}{\langle N \rangle}$, and thus the criterion is naturally satisfied (also see \aref{restart-detrimental} for counter cases). Note from \fref{Non biased RW} and the inset of \fref{Non biased RW-criterion}b that
$\langle N_R \rangle$ has a minimum at the optimal probability $p^*$. This value can furthermore be conferred from the root equation (\ref{optimal p}). 
Finally, the probability mass function of the first passage time namely $P_{N_R}(n)$ is obtained analytically from \eref{PGF PMF} and \eref{PGF geometric} by using $G_N(z)$ and plotted in \fref{Non biased RW-criterion}b against the data obtained from the numerical simulations. The inset of \fref{Non biased RW-criterion}b corroborates with the parameter $n_0=3$ (taken from \fref{Non biased RW-criterion}a) which guarantees a reduction in $\langle N_R \rangle$ in the presence of restart. Here, we comment on the asymptotic large $n$-form of $P_{N_R}(n)$. It is intuitive to understand that a successful first passage event would occur after a few restart events that brings back the walker to its initial position. Thus, if $Y$ denotes the number of restart events until a first passage then it simply conforms to a geometric distribution. Moreover, if $\tilde{Y}$ is the time until the first passage under $Y$, then the average of this random time is given by $\langle N_R \rangle$. It is then trivial to show that Prob$(\tilde{Y}>t) \to e^{-t/\langle N_R \rangle}$ as $t \to \infty$ \cite{Mor}, which essentially implies that $P_{N_R}(n)$ is asymptotically exponential in $n$. We demonstrate this asymptotic form in \fref{scaling}.

\begin{figure}[b]
\includegraphics[width=8cm,height=4.cm]{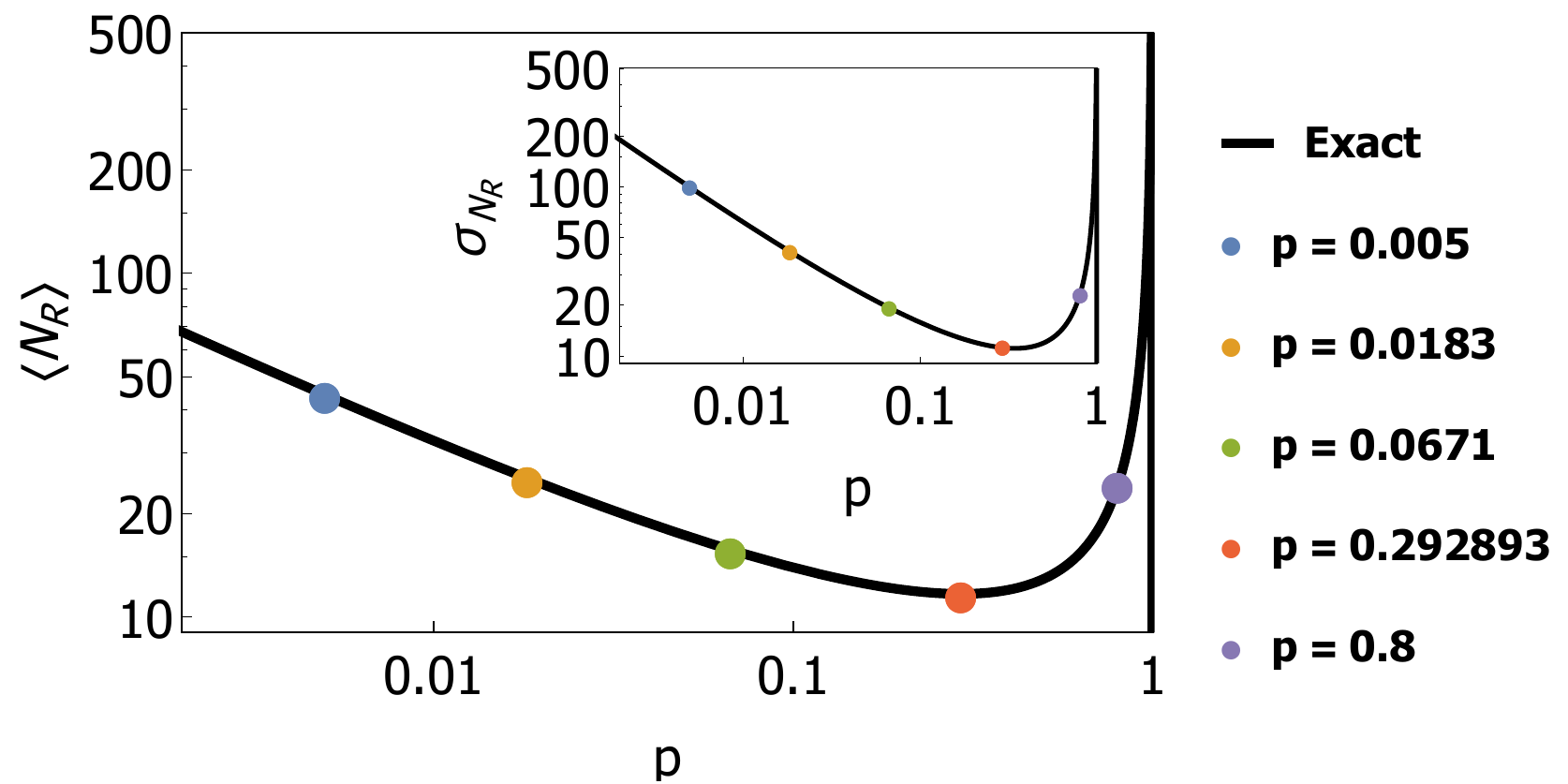}
\caption{Mean and standard deviation for the simple RW on the semi-infinite line with resetting. Parameters: left boundary $ = -2$, initial position $n_0 = 0$, and $q=1$. The right boundary is set to infinity.}
\label{Simple random walk mean and std}
\end{figure}

It is only imperative to revisit the simple random walker (with $q=1$ and starting at the origin) in the presence of only one absorbing boundary located at $x$. In this case, the generating function for the underlying process can be easily obtained following steps from \cite{LRW-5} and this reads 
\begin{align}
    G_N(z) =\left(\frac{1-\sqrt{1-z^2}}{z}\right)^{|x|},\quad x\neq 0. \label{Simple random walk first passage}
\end{align}
In fact, it is known that for large $n$, the first passage time density has a power law tail $n^{-3/2}$ \cite{LRW-1,LRW-2,LRW-3,LRW-4}, which is similar to the L\'{e}vy-Smirnov distribution for the first passage time of a Brownian walker in one dimension. This power law trivially leads to a diverging mean first passage time for a RW. Naturally, it is only expected that restart will always expedite the first passage time as was shown in the classic diffusion problem with stochastic resetting \cite{Restart1}. For completeness, in this case, we compute the mean $\langle N_R \rangle$ and the fluctuations $\sigma_{N_R}$ as a function of the restart probability $p$ from the generating function using Eqs. (\ref{mean FPUR geometric}) and (\ref{second moment FPUR geometric}). The theoretical results are in excellent agreement with the numerical simulations of the walker in the semi-infinite geometry (see \fref{Simple random walk mean and std}). We end this discussion by referring the readers to \aref{Sharp-numerics} where we have presented the results for the RW subject to sharp resetting.

\begin{figure}[t]
\includegraphics[width=8.5cm,height=4.25cm]{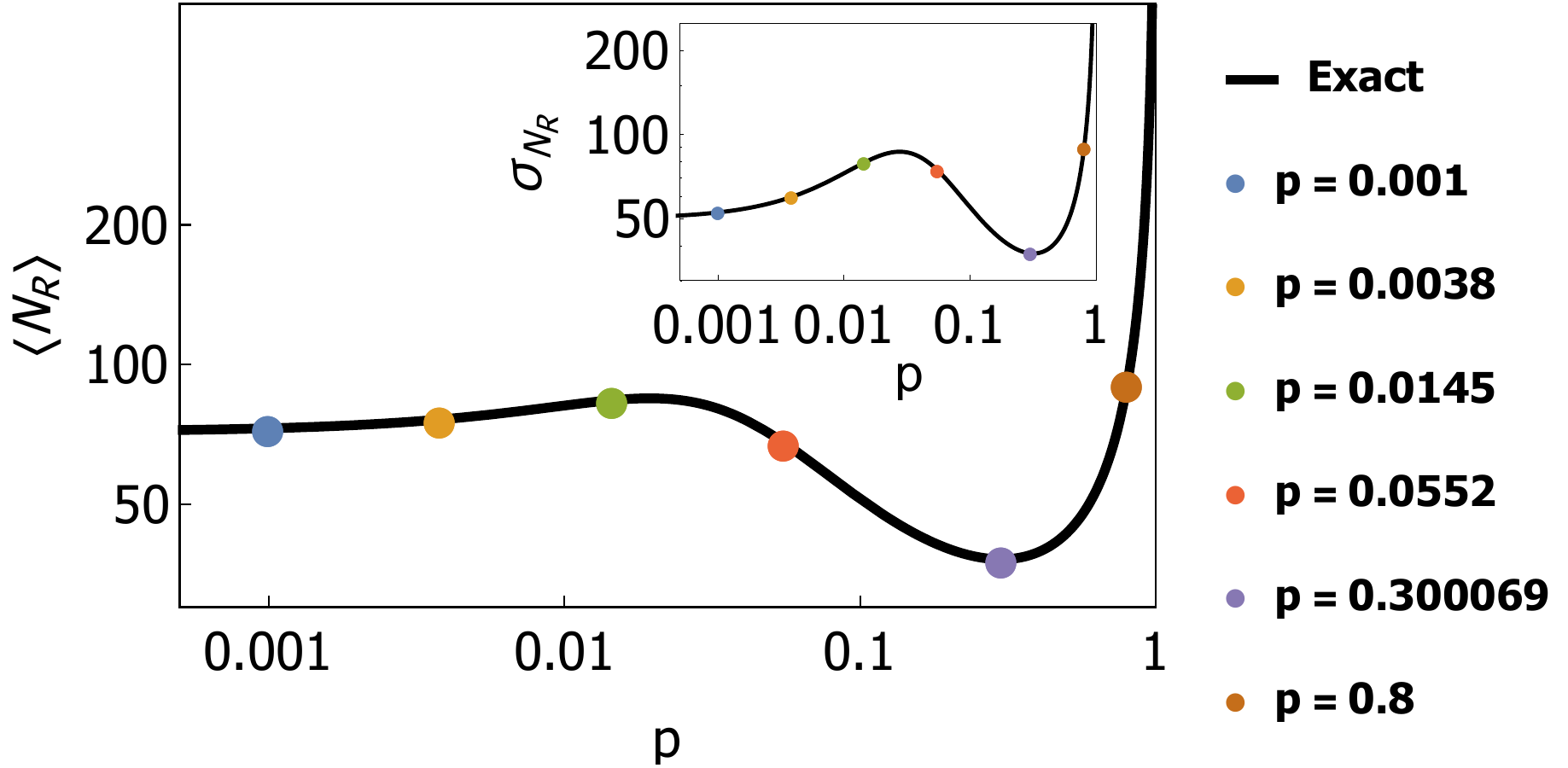}
\caption{Comparison between exact and simulation results for mean and standard deviation for a biased RW in confinement subject to restart conducted at a probability $p$. Parameters for the current set-up: left boundary $ = 1$, right boundary $ = 25$, initial position  $n_0 = 3$, diffusive parameter $q = 0.7$, and bias $g = 0.3$.}
\label{ biased RW }
\end{figure}

\begin{figure*}[t]
\includegraphics[width=8.2cm,height=4.3cm]{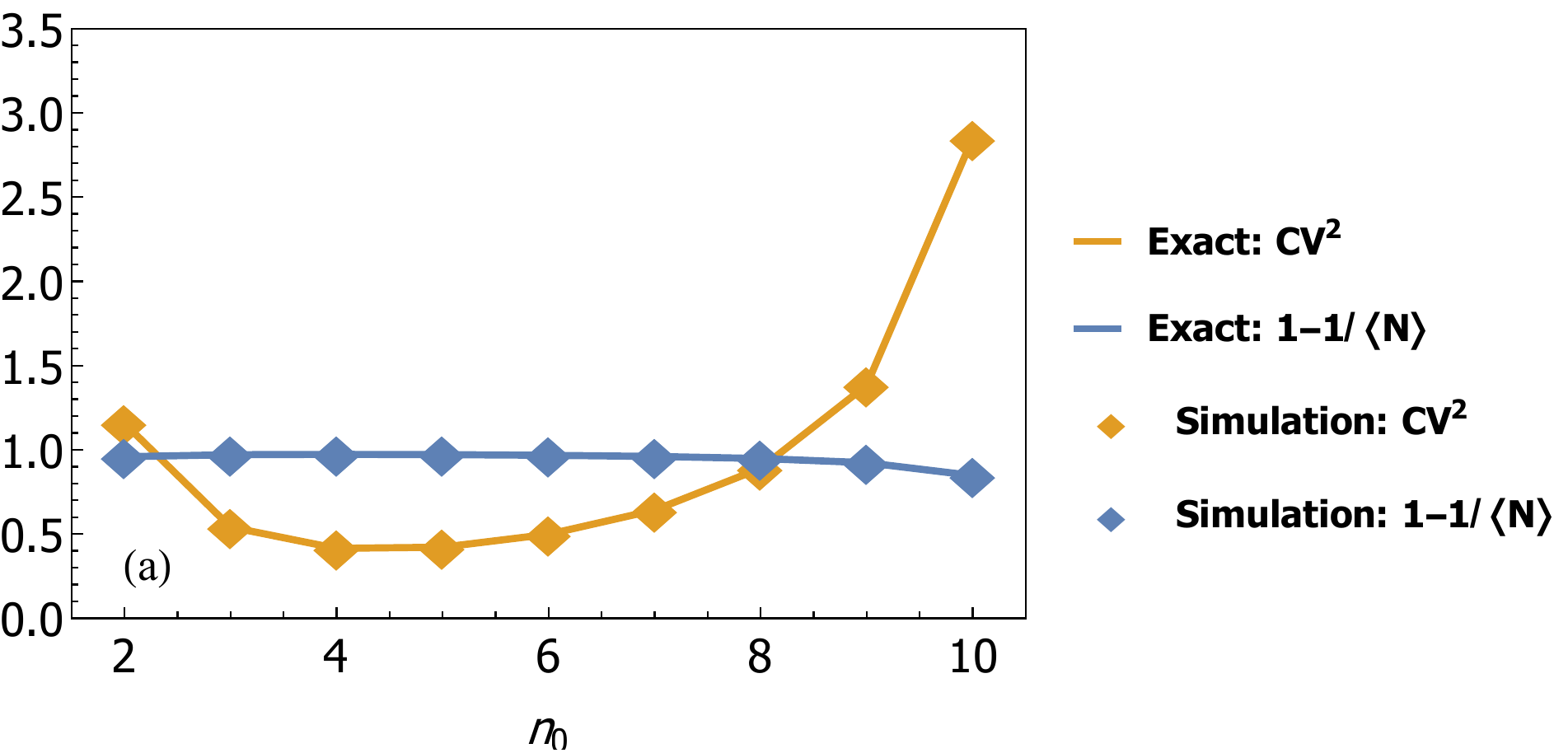}
\includegraphics[width=8.2cm,height=4.2cm]{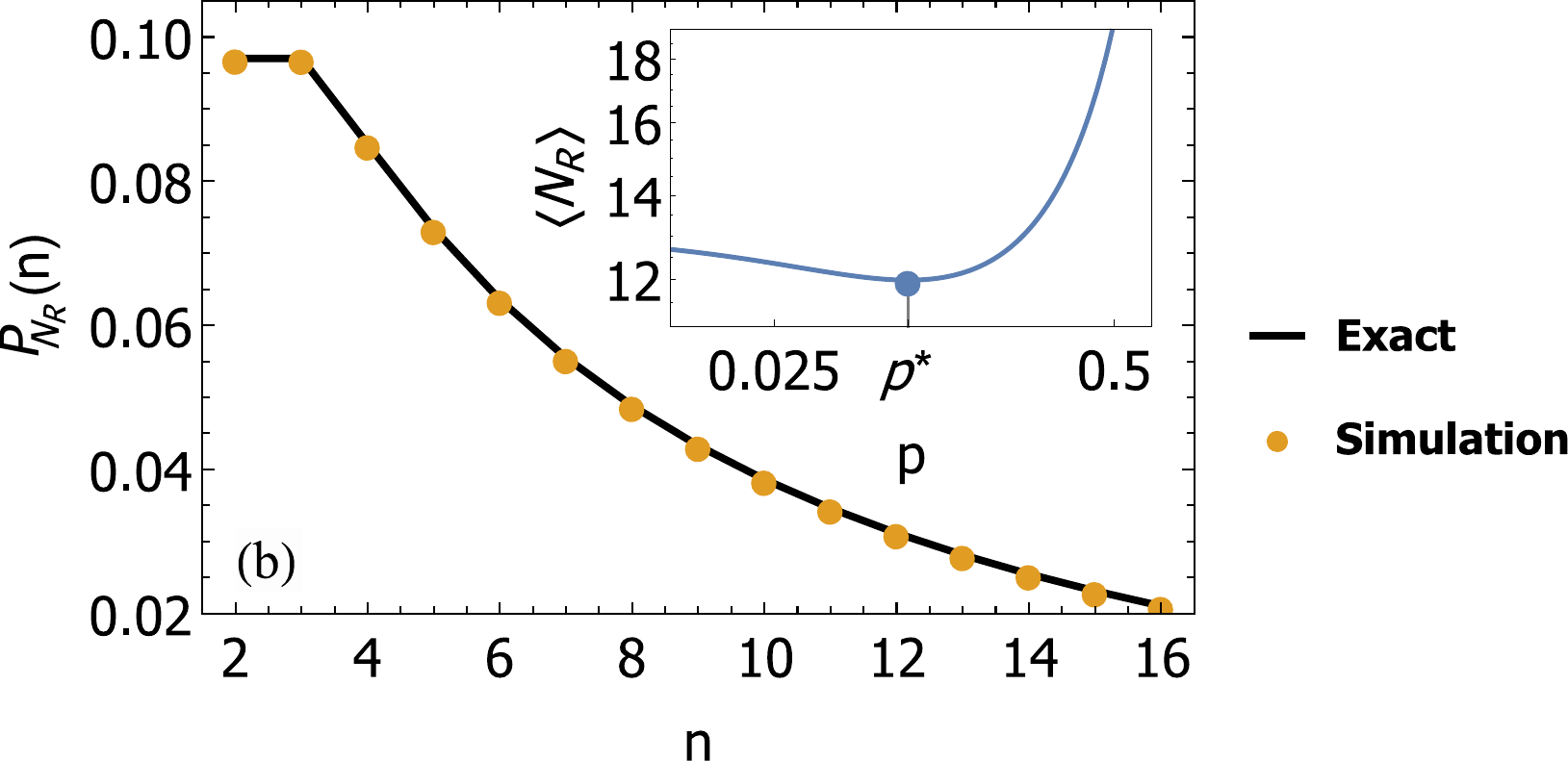}
\caption{Panel (a): demonstration of the restart criterion (\ref{CV crierion}) for the biased RW in confinement. We plotted the LHS and RHS of the criterion as a function of the initial position $n_0$. Panel (b): we choose $n_0=9$ from the panel (a) and then plot $P_{N_R}(n)$ for this given parameter value. In the inset, we plot $\langle N_R \rangle$ to demonstrate that indeed restart is able to reduce the mean completion time. The optimal probability $p^*=0.0813..$ is found to match exactly with the theoretical prediction. Parameters fixed for the set-up: left boundary $ = 1$, right boundary $ = 11$
and $q = 0.5,~g = 0.3$.}
\label{biased RW CV criteria}
\end{figure*}

\subsection{Biased random walks in 1D}
\label{example-2}

We now consider a biased random walker in the presence of restart. The underlying problem was recently studied by Sarvaharman \& Giuggioli in \cite{Luca2} generalizing the results obtained in \cite{Luca1}. Biased lattice RWs are quite important since their applications include cell migration due to concentration gradients in biology
(chemotaxis) \cite{chemo1,chemo2}, drifting bacteria by light modulation (phototaxis) \cite{photo}, or upwards movement of single-celled algae in response to gravity (gravitaxis) \cite{gravity}. Moreover, the model has been employed to study wireless sensor networks \cite{network}, model driven tracer particles \cite{tracer}. Although seemingly there is a lot of interests, an attempt to derive exact expressions for first-passage statistics for biased lattice RW in confined space has been extremely limited. The work by Sarvaharman \& Giuggioli in \cite{Luca2} develops a general framework that allows to derive analytically various transport quantities in arbitrary dimensions and
arbitrary boundary conditions for biased lattice random walk. Since the problem with restart can be elegantly mapped to the problem without restart, we use some of the results obtained in \cite{Luca2}. We first recall the model for brevity.

We start by considering the dynamics of a random walker with bias on a 1D confined lattice with two absorbing points at the end.
Here, the strength and direction of the bias is described by the parameter $g$. Thus, we assume that the probability of jumping to the neighboring site on the
left is given by $\frac{q}{2}(1 - g)$, and the probability of jumping to right is then given by $\frac{q}{2}(1 + g)$, where recall that $q$ is the diffusivity parameter. Thus, probability of staying in the same site is given by $1-q$. When $g = 0$, the movement is diffusive, whereas the cases $g = 1$ and $g = -1$ are, respectively, the ballistic limit to the right and left sites.
For this RW, the dynamics is governed by the following evolution equation for the occupation probability
\bea
  \mathbb{P}(m, n+1) &= (1 - q) \mathbb{P}(m, n) + \frac{q}{2}(1 - g) \mathbb{P}(m - 1, n) \nonumber
  \\ &+ \frac{q}{2}(1 + g) \mathbb{P}(m + 1, n).
\label{bias mater eq}
\eea
For a walker performing the RW described in \eref{bias mater eq}, in a bounded interval $(1,N)$ and starting at $1<n_0<N$, where $m = 1$ and $m = N$ are absorbing boundaries (with $\mathbb{P}(1,n)=\mathbb{P}(N,n)=0$), the propagator is given by \cite{Luca2}
\bea 
  \mathbb{P}(m, n) = 
  \sum_{k = 1}^{N-2} h_{k} \left(m, n_0 \right) \left[1 + s_k\right]^{n},\label{Biased propagator}
\eea 
where $s_k$ and $h_k$ are given by \cite{Luca2}
\bea
s_k &=& \frac{q}{\eta}\cos{\left( \frac{k \pi}{N - 1} \right)} - q, \nn \\ \label{s function}
h_{k}(m,n_0) &=& \frac{2f^{\frac{m - n_0}{2}}\sin \left[ \left( \frac{m-1}{N-1} \right ) k\pi \right] \sin\left[ \left(\frac{n_0-1}{N-1}\right)k\pi \right]}{N-1}, \label{h function}
\eea 
and
\bea
f = \frac{1 + g}{1 - g},
\quad \eta = \frac{1 + f}{2 \sqrt{f}}. \label{f and eta}
\eea 
Again, here we need only the generating function $G_N(z)$ for the underlying first passage time density $P_N(n)$ to make use of our renewal formulas. To compute this, we note that, by definition, 
\bea
P_N(n) = \mathcal{S}_{n_{0}}\left(n-1\right)-\mathcal{S}_{n_{0}}\left(n\right), \label{FP from survival}
\eea 
where $\mathcal{S}_{n_{0}}\left(n\right)$ 
is the survival probability i.e., the probability that the walker survives upto the time step $n$ starting from $n_0$ (without restart) and is given by $\mathcal{S}_{n_{0}}\left(n\right)=\sum_{m=1}^{N}\mathbb{P}\left(m,n\right)$, where $\mathbb{P}\left(m,n\right)$ is given by \eref{Biased propagator} \cite{LRW-1}. Replacing $\mathcal{S}_{n_{0}}\left(n\right)$ in \eref{FP from survival} gives an exact expression for $P_N(n)$ (also see \fref{Underlying-FPT}b for a numerical verification). Next, to obtain the generating function $G_N(z)$, we multiply $z^n$ on the both sides of \eref{FP from survival} and sum over $n$. From the resulting expression for $G_N(z)$, the mean $\langle N_R \rangle$ and the fluctuations $\sigma_{N_R}$ are computed using Eqs. (\ref{mean FPUR geometric}) and (\ref{second moment FPUR geometric}) simply by replacing $z \to 1-p$. We have plotted these expressions as a function of restart probability $p$ in \fref{ biased RW }. Similar to the symmetric RW, we now investigate the criterion (\ref{CV crierion}) in \fref{biased RW CV criteria}a. We plot $CV^2$ and $1-\frac{1}{\langle N \rangle}$ as a function of the initial position $n_0$. We choose an $n_0$ for which restart is beneficial and corroborate with \fref{biased RW CV criteria}b to plot the probability mass function for the completion time of the restarted process. From the inset, it is clear that restart lowered the completion time and thus implying the existence of an optimal restart probability $p^*$. We compare this value with that obtained from the theory (\eref{optimal p}) to get an exact match. The large $n$ asymptotic form of $P_{N_R}(n)$ for the biased RW case has also been verified in \fref{scaling}. Finally, we refer to \aref{Sharp-numerics} for the results in the presence of sharp restart.

\section{Conclusions}
\label{conclusions}
In summary, we have studied first passage under restart when both underlying and restart processes are spatially and temporally discrete. Although there exists a plethora of works dedicated on Brownian walkers subject to restart, results on a discrete time and space random walker are only handful and problem specific. Taking advantage of the renewal properties, we have derived working formulas for the mean and higher order moments of the completion time under restart. Importantly, the formulation is quite general and can be used as a platform to extend our findings to
a wide range of other discrete stochastic motions and restart time distributions. In particular, we have shown that when restart is geometrically distributed, there is a sufficient criterion which determines when restart is beneficial. We apply the theory in two paradigmatic set-ups namely the canonical 1D lattice random walker (symmetric and biased) in the presence of absorbing boundaries at two given end points. Using them as underlying first passage time processes, we analyze the effect of restart in full details. We stress the fact that it is also possible to compute the density function of the first passage time under restart at any time unlike the continuous cases where only large time asymptotics are tractable. Our analytical results provide an excellent agreement with all the simulation results. There are many extensions possible within this current set-up. Our method can be employed directly to understand the effects of restart on the search or first passage time dynamics of confined lattice RWs in arbitrary dimensions. It will be interesting to study the effects of sticky or mixed boundaries for RW models in the presence of restart with an overhead time \cite{HRS,overhead-1,overhead-1-1,overhead-2} or a refractory period \cite{overhead-3}. Random walks in a spatially non-homogeneous condition under restart is also another challenging direction. Consider a simple RW on a lattice and start diluting the lattice. By this we
mean that some fraction of the lattice sites is removed, i.e., is declared inaccessible for the walker. This is often known as random walks on a percolation structure \cite{LRW-5}. It would be interesting to study how the probability of finding a finite cluster among all finite clusters scales in the presence of restart.

\begin{acknowledgments}
We thank Shlomi Reuveni and Yuval Schehr for many fruitful discussions. Arnab Pal gratefully acknowledges support from the Raymond and  Beverly  Sackler  Post-Doctoral  Scholarship and the Ratner Center for Single Molecule Science at  Tel-Aviv University. 
\end{acknowledgments}

\appendix

\section{Derivation for the generating function}
\label{derive-G}
In this section, we present the derivation of the PGF of the first passage process under generic restart. To do so, we note that 
\bea 
G_{N_{R}}(z) &=&\text{Pr}\left(N\leq R\right)\left\langle Z^{N_{R}}|N\leq R\right\rangle \nn \\ &+& \text{Pr}\left(N > R\right)\left\langle Z^{N_{R}}|N> R\right\rangle, \label{PGF derivation 1} 
\eea 
which gives
\bea 
G_{N_{R}}(z)&=&\text{Pr}\left(N \leq R\right)\left\langle Z^{\left\{ N_{R}|N \leq R\right\} }\right\rangle \nn \\ &+& \text{Pr}\left(N > R\right)\left\langle Z^{\left\{ N_{R}|N > R\right\} }\right\rangle. \label{PGF derivation 2}
\eea 
Recall the random variables defined in the main text
\begin{align}
N_{min}&=\left\{ N_{R}|N \leq R\right\} =\left\{ N|N \leq R\right\} =\left\{ N|N=min\left(R,N\right)\right\}, \label{Tmin} \\
R_{min}&=\left\{ N_R|N> R\right\} =\left\{ R|N> R\right\} =\left\{ R|R=min\left(R,N\right)\right\}, \label{Rmin}
\end{align}
where $min\left(R,N\right)$ is the minimum of $N$ and $R$. Note that
\begin{align}
\left\{ N_{R}|N> R\right\} &=\left\{ R+N'_{R}|N> R\right\} \nn \\ 
&=\left\{ R|R=min\left(R,N\right)\right\} +N'_{R}=R_{min}+N'_{R}, \label{Rmin relation}
\end{align}
where in the second transition in \eref{Rmin relation} we have further used the fact that $N'_{R}$ is an independent and identically distributed copy of $N_{R}$ and hence independent of both $R$ and $N$. We thus have
\begin{align}
G_{N_{R}}(z)&=\text{Pr}\left(N \leq R\right)\left\langle Z^{N_{min}}\right\rangle +\text{Pr}\left(N > R\right)\left\langle Z^{R_{min}+T'_{R}}\right\rangle \nn \\
&=\text{Pr}\left(N \leq R\right)G_{N_{min}}(z)+\text{Pr}\left(N > R\right)G_{R_{min}}(z)G_{N_{R}}(z), \label{PGF derivation 3}
\end{align}
where in the last step we have again used the fact that $N'_{R}$ is an independent
and identically distributed copy of $N_{R}$. Rearranging \eref{PGF derivation 3} we have
\bea 
G_{N_{R}}(z)&=&\frac{\text{Pr}\left(N \leq R\right)G_{N_{min}}(z)}{1-\text{Pr}\left(N > R\right)G_{R_{min}}(z)} \nn \\
&=&\frac{\left(1-\text{Pr}\left(N > R\right) \right)G_{N_{min}}(z)}{1-\text{Pr}\left(N > R\right)G_{R_{min}}(z)}, \label{PGF final result-A}
\eea 
which is \eref{PGF final result} in the main text.

\begin{figure*}[t]
\includegraphics[width=8.2cm,height=4.3cm]{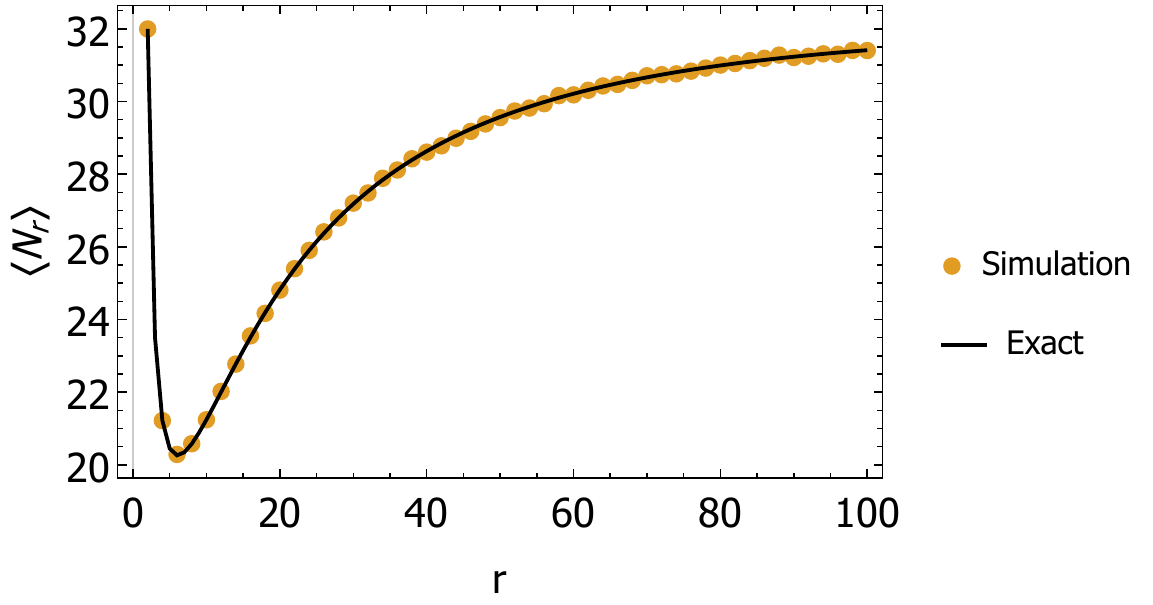}
\includegraphics[width=8.2cm,height=4.2cm]{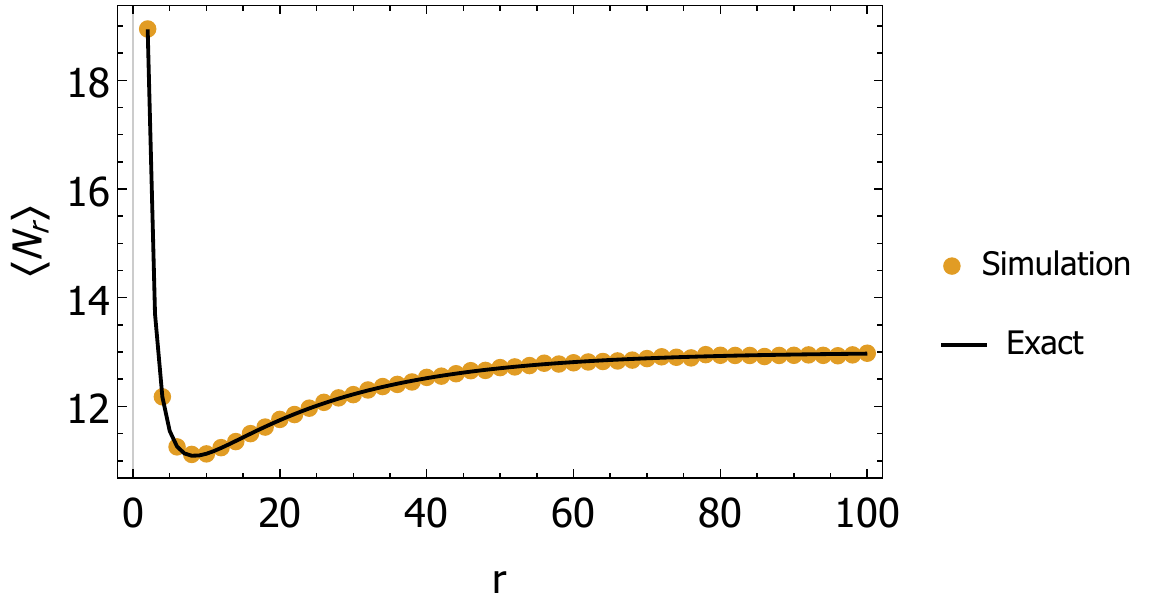}
\caption{Mean first passage time $\langle N_r \rangle$ under sharp restart has been plotted against the restart step $r$. Left panel: symmetric RW. Parameters: left boundary $ = 1$, right boundary $ = 11$ and $n_0 = 3, q = 0.5$. Right panel: biased RW. Parameters: left boundary $ = 1$, right boundary $ = 11$ and $n_0 = 3, ~q = 0.5, ~g = -0.3$. In both cases, the mean times saturate to their underlying values as $r \gg 1$.}
\label{Sharp-plots}
\end{figure*}

\section{Derivations for first passage step under geometric restart}
\label{derivation-geometric-appendix}
In this section, we present the derivations for Eqs. (\ref{mean FPUR geometric}) and (\ref{PGF geometric}).
We start the derivation by noting that under geometrically distributed resetting, $\text{Pr}\left(N\leq R\right)$ is given by
\bea 
\text{Pr}\left(N\leq R\right)&=&\sum_{n=0}^{\infty}P_{N}(n)\sum_{m=n}^{\infty}P_{R}(m) \nn \\
&=&\sum_{n=0}^{\infty}P_{N}(n)\sum_{m=n}^{\infty}(1-p)^{m} p \nn \\
&=&\sum_{n=0}^{\infty}P_{N}(n) (1-p)^{n} \nn \\
&=&G_N(1-p), \label{mean denominator geometric} \eea 
where in the last step we utilized the definition of the PGF of random variable $X$, given in \eref{PGF def}.
Using \eref{Mean min formula}, one can also compute $\langle min\left(N,R\right)\rangle$ for the case of geometrically distributed restart
\bea 
\left\langle min\left(N,R\right)\right\rangle&=& \sum_{n=0}^{\infty} \left(\sum_{k=n+1}^{\infty}P_N\left(k\right)\right) \left(\sum_{m=n+1}^{\infty}P_R\left(m\right)\right) \nn \\
&=& \sum_{n=0}^{\infty} \left(\sum_{k=n+1}^{\infty}P_N\left(k\right)\right) \sum_{m=n+1}^{\infty}(1-p)^m p \nn \\ 
&=& \sum_{n=0}^{\infty} \sum_{k=n+1}^{\infty}P_N\left(k\right) (1-p)^{n+1}  \nn \\
&=& \sum_{k=1}^{\infty} \sum_{n=0}^{k-1}P_N\left(k\right) (1-p)^{n+1}  \nn \\
&=& \sum_{k=1}^{\infty} P_N\left(k\right) \sum_{n=0}^{k-1} (1-p)^{n+1}  \nn \\
&=& \sum_{k=1}^{\infty} P_N\left(k\right) \left ( \frac {1 - p} {p} + \frac {(p - 1) (1 - p)^k} {p} \right)  \nn \\
&=& \sum_{k=0}^{\infty} P_N\left(k\right) \left ( \frac {1 - p} {p} - \frac {(1 - p) (1 - p)^k} {p} \right)  \nn \\
&=& \frac {1 - p} {p}\left(1-\sum_{k=0}^{\infty} P_N\left(k\right)(1 - p)^k  \right) \nn \\
&=& \frac {1 - p} {p}\left(1-G_N(1-p)  \right), \label{Mean min geometric}
\eea 
where in the second to last step we once again utilize the definition of PGF. Equation (\ref{mean FPUR geometric}) is obtained by substituting Eqs. (\ref{mean denominator geometric}) and (\ref{Mean min geometric}) into \eref{mean FPUR}.

We now turn to the derivation of the PGF of the restarted first passage process $N_R$. We start by deriving $G_{N_{min}}(z)$ and $G_{N_{R}}(z)$. We substitute the PMF of the geometric distribution into Eqs. (\ref{Nmin mass-main}) and (\ref{Rmin mass-main}) to find
\bea
G_{N_{min}}(z)
&=& \sum_{n=0}^{\infty} P_{N}(n)\frac{\sum_{m=n}^{\infty}P_{R}(m)}{\text{Pr}\left(N \leq R\right)} z^n \nn \\
&=& \sum_{n=0}^{\infty} P_{N}(n)\frac{\sum_{m=n}^{\infty}(1-p)^m p}{G_N(1-p)} z^n\nn \\
&=& \sum_{n=0}^{\infty} P_{N}(n)\frac{(1-p)^{n}}{G_N(1-p)} z^n \nn \\
&=& \frac{1}{G_N(1-p)}\sum_{n=0}^{\infty} P_{N}(n)(1-p)^{n} z^n \nn \\
&=& \frac{G_N(z(1-p))}{G_N(1-p)}, \label{GNmin geometric}
\eea
and
\bea
G_{R_{min}}(z)&=& \sum_{n=0}^{\infty} P_{R}(n)\frac{\sum_{m=n+1}^{\infty}P_{N}(m)}{\text{Pr}\left(N > R\right)} z^n \nn \\
&=& \sum_{n=0}^{\infty} (1-p)^n p  \frac{\sum_{m=n+1}^{\infty}P_{N}(m)}{1-\text{Pr}\left(N \leq R\right)} z^n  \nn \\
&=& \frac{1}{1-G_N(1-p)} \sum_{n=0}^{\infty}\sum_{m=n+1}^{\infty} (1-p)^n p z^n P_{N}(m)  \nn \\
&=& \frac{1}{1-G_N(1-p)} \sum_{m=1}^{\infty}\sum_{n=0}^{m-1} (1-p)^n p z^n P_{N}(m)  \nn \\
&=& \frac{1}{1-G_N(1-p)} \sum_{m=1}^{\infty} P_{N}(m) \sum_{n=0}^{m-1} (1-p)^n p z^n \nn \\
&=& \frac{1}{1-G_N(1-p)} \sum_{m=1}^{\infty} P_{N}(m) \left( \frac{p \left(((1-p) z)^m-1\right)}{(1-p)
   z-1}\right) \nn \\
&=& \frac{p\left( \sum_{m=1}^{\infty} P_{N}(m)((1-p) z)^m-\sum_{m=1}^{\infty} P_{N}(m) \right)}{((1-p)
   z-1) \left(1-G_N(1-p)\right)} \nn \\
&=& \frac{p\left( \sum_{m=0}^{\infty} P_{N}(m)((1-p) z)^m-\sum_{m=0}^{\infty} P_{N}(m) \right)}{((1-p)
   z-1) \left(1-G_N(1-p)\right)} \nn \\
&=& \frac{p\left( G_N(z(1-p))-1 \right)}{((1-p)
   z-1) \left(1-G_N(1-p)\right)}. \label{GRmin geometric}
\eea
Equation (\ref{PGF geometric}) is obtained by substituting Eqs. (\ref{mean denominator geometric}), (\ref{GNmin geometric}) and (\ref{GRmin geometric}) into \eref{PGF final result}.

Taking the derivative of \eref{PGF geometric}, $z\rightarrow 1^-$, one can recover the mean first passage time under geometric restart given in \eref{mean FPUR geometric}
\bea 
\langle N_R \rangle &=&  z \frac{\partial G_{N_R}(z)}{\partial z} \Big|_{z=1^{-}} \nn \\
&=& \scriptstyle{z\frac{(1-p) \left((1-p) (z-1) ((1-p) z-1)
   G'_N((1-p) z)-p G_N((1-p)
   z)^2+p G_N((1-p) z)\right)}{((1-p)
   (z-1)-p G_N((1-p) z))^2}\Big|_{z=1^{-}}} \nn \\
&=& \frac {1 - p} {p} \frac{1-G_N(1-p)}{G_N(1-p)}, \label{mean FPUR geometric 2}
\eea 
which is identical to the result obtained in \eref{mean FPUR geometric}.

\section{Derivations for first passage step under sharp restart}
\label{Sharp-numerics}
In this section, we sketch the steps to derive \eref{PGF sharp-main}.
As for the geometric resetting case, we start the derivation by obtaining $G_{N_{min}}(z)$ and $G_{N_{R}}(z)$ using Eqs. (\ref{Nmin mass-main}) and (\ref{Rmin mass-main})
\bea
G_{N_{min}}(z)&=& \sum_{n=0}^{\infty} P_{N}(n)\frac{\sum_{m=n}^{\infty}P_{R}(m)}{\text{Pr}\left(N \leq R\right)} z^n \nn \\
&=& \sum_{n=0}^{\infty} P_{N}(n)\frac{\sum_{m=n}^{\infty}\delta_{m,r}}{\text{Pr}\left(N \leq r\right)} z^n\nn \\
&=& \frac{1}{\text{Pr}\left(N \leq r\right)} \sum_{n=0}^{r} P_{N}(n) z^n. \label{GNmin sharp}
\eea 

\begin{figure}[b]
\includegraphics[width=7.5cm,height=5cm]{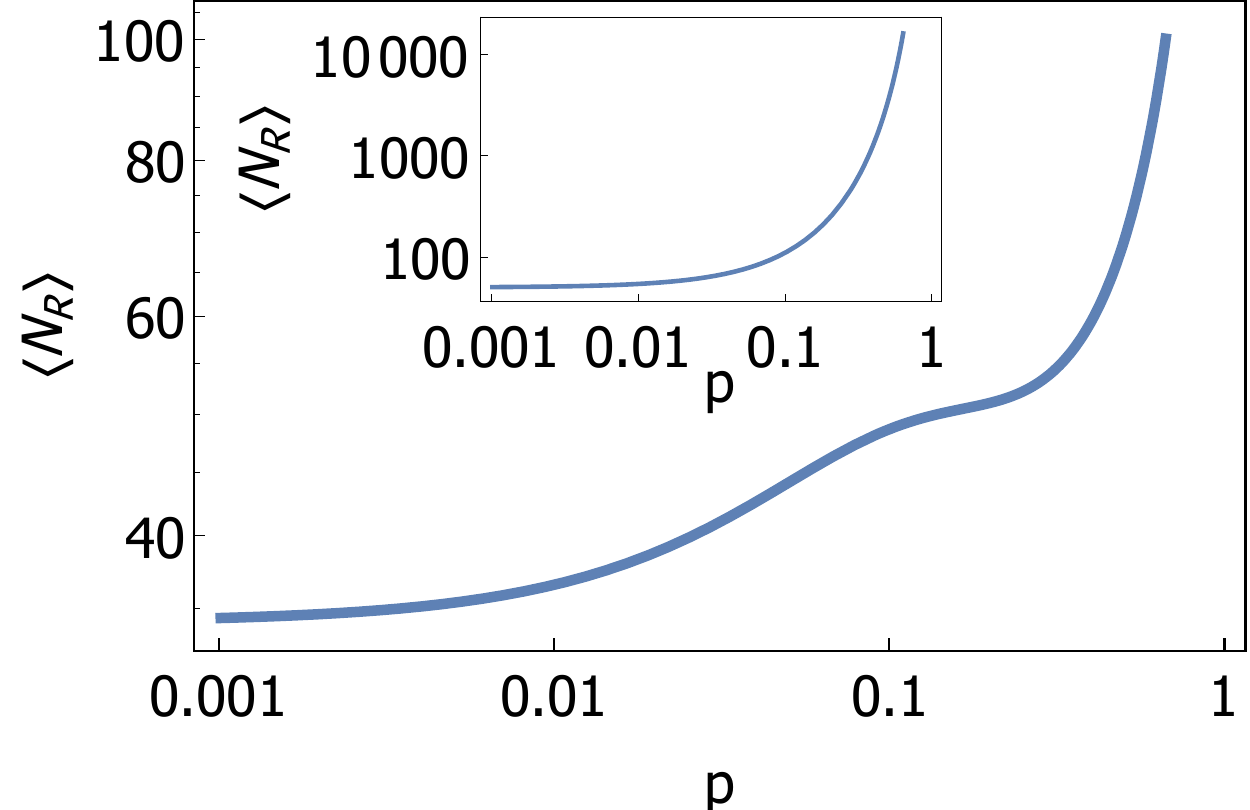}
\caption{Mean first passage time under geometric restart for the case when restart is detrimental. We refer to \fref{Non biased RW-criterion}a and \fref{biased RW CV criteria}a for the simple and biased RWs respectively. In particular, we choose those values of $n_0$ for which the criterion (\ref{CV crierion}) is violated. It can be seen that adding restart only increases $\langle N_R \rangle$. Parameters set for the biased RW: left boundary $ = 1$, right boundary $ = 11$, and $n_0 = 3, ~q = 0.5, ~ g = 0.3$ and for the symmetric RW (inset): left boundary $ = 1$, right boundary $ = 11$ and $n_0=6,~q=0.5$.}
\label{geometric-detrimental}
\end{figure}

Moreover, a similar exercise gives
\bea
G_{R_{min}}(z)&=& \sum_{n=0}^{\infty} P_{R}(n)\frac{\sum_{m=n+1}^{\infty}P_{N}(m)}{\text{Pr}\left(N > R\right)} z^n \nn \\
&=& \sum_{n=0}^{\infty} \delta_{n,r}  \frac{\sum_{m=n+1}^{\infty}P_{N}(m)}{\text{Pr}\left(N > r\right)} z^n  \nn \\
&=& \frac{\sum_{m=r+1}^{\infty}P_{N}(m)}{\text{Pr}\left(N > r\right)} z^r \nn \\
&=& \frac{\text{Pr}\left(N > r\right)}{\text{Pr}\left(N > r\right)} z^r = z^r. \label{GRmin sharp}
\eea 
Equation (\ref{PGF sharp-main}) is then obtained by substituting Eqs (\ref{GNmin sharp}) and (\ref{GRmin sharp}) into \eref{PGF final result} i.e.,
\begin{align}
G_{N_{R}}(z)&=\frac{\text{Pr}\left(N \leq R\right)G_{N_{min}}(z)}{1-\text{Pr}\left(N > R\right)G_{R_{min}}(z)} \nn \\
&= \frac{\text{Pr}\left(N \leq r\right)\frac{1}{\text{Pr}\left(N \leq r\right)} \sum_{n=0}^{r} P_{N}(n) z^n}{1-\text{Pr}\left(N > r\right)z^r} \nn \\
&= \frac{\sum_{n=0}^{r} P_{N}(n) z^n}{1-\text{Pr}\left(N > r\right)z^r}. \label{PGF sharp}
\end{align}
In \fref{Sharp-plots}, we plot the mean completion time $\langle N_r \rangle$ as a function of restart step $r$ for the symmetric and biased RW. For small $r$, restart events are quite frequent and thus the completion times are large. On the other hand, for $r \gg 1$, restart rarely occurs and we reach to the limit of the underlying process.

\vspace{0.35cm}

\section{Cases when restart is detrimental}
\label{restart-detrimental}
In \sref{CV-criterion-geometric} (main text), we have derived a sufficient criterion for restart to be beneficial (see \eref{CV crierion}). We presented examples of RW with restart to demonstrate the criterion. Here, we present the same examples, but show that restart can be detrimental on the violation of the criterion (\ref{CV crierion}). Recall that for a biased RW, \fref{biased RW CV criteria}a pictorially depicts the criterion as a function of $n_0$. Thus, we take a value of $n_0$ (say $n_0=3$) for which the criterion is not satisfied. Restart is expected only to prolong the completion in such case. This is depicted in \fref{geometric-detrimental}. A similar analysis is also made for the symmetric RW following \fref{Non biased RW-criterion}a and the resulting plot for the mean completion time as a function of restart probability $p$ is shown in the inset of \fref{geometric-detrimental}. As expected, mean time increases as $p$ varies showcasing another example of restart being detrimental.

\end{document}